\documentclass[journal,10pt]{IEEEtran}
\usepackage{amsfonts}
\usepackage{amsmath}
\usepackage{amssymb}
\usepackage{bm}
\usepackage{epsfig}
\usepackage{epstopdf}
\usepackage{graphicx}
\usepackage[font=small,labelsep=period]{caption}
\usepackage{cases}
\usepackage{cite}
\usepackage{color}
\usepackage{multicol}
\usepackage{exscale}
\usepackage{float}
\usepackage{graphics}
\usepackage{graphicx}
\usepackage{latexsym}
\usepackage{subfigure}
\usepackage{latexsym}
\usepackage{stfloats}
\usepackage{exscale}
\usepackage{relsize}
\usepackage[ruled,linesnumbered]{algorithm2e}
\usepackage{mathtools}
\usepackage{soul}
\begin{document}
\title{Two-Timescale Transmission Design for Wireless Communication Systems Aided by Active RIS}
\author{
        Zhangjie~Peng,
        Tianshu~Li,
        Cunhua~Pan,~\IEEEmembership{Member,~IEEE},
        \\
        Xianfu Lei ,~\IEEEmembership{Member,~IEEE},
        and~Shi~Jin,~\IEEEmembership{Senior~Member,~IEEE}\vspace{-0.2cm}
\vspace{-0.15cm}
\thanks{Z. Peng is with the College of Information, Mechanical and Electrical Engineering, Shanghai Normal University, Shanghai 200234, China $( \text{e-mail: pengzhangjie@shnu.edu.cn} )$.
\emph{(Corresponding authors: Cunhua Pan.)}}
\thanks{T. Li and X. Lei are with the School of Information Science and Technology, Southwest Jiaotong University, Chengdu 610031, China (2022300282@my.swjtu.edu.cn; xflei@home.swjtu.edu.cn).}
 \thanks{C. Pan and S. Jin are with the National Mobile Communications Research Laboratory, Southeast University, Nanjing 210096, China.
 (e-mail: cpan@seu.edu.cn; jinshi@seu.edu.cn).}
 \vspace{-0.9cm}
}

\maketitle

\IEEEoverridecommandlockouts

\newtheorem{lemma}{Lemma}
\newtheorem{theorem}{Theorem}
\newtheorem{remark}{Remark}
\newtheorem{corollary}{Corollary}
\newtheorem{proposition}{Proposition}

\begin{abstract}
This paper considers an active reconfigurable intelligent surface (RIS)-aided communication system, where an $M$-antenna base station (BS) transmits data symbols to a single-antenna user via an $N$-element active RIS.
We use two-timescale channel state information (CSI) in our system,
so that the channel estimation overhead and feedback overhead can be decreased dramatically.
For both the uplink and downlink systems, we derive the closed-form approximate expressions of the achievable rates (ARs) and propose the phase shift optimization schemes.
In addition, we compare the performance of the active RIS system with that of the passive RIS system. The numerical results show that the active RIS system achieves a larger AR than the passive RIS system.

\end{abstract}

\section{Introduction }

Reconfigurable intelligent surface (RIS) has attracted extensive research attention, which has been envisioned  as one of the potential key technologies in 6G communication systems \cite{8741198}.
In particular, an RIS is equipped with a large number of passive and low-cost reflecting elements, each of which can induce an independent phase shift on the incident signals to enhance the received signal.
Due to its passive nature, the RIS possesses appealing features of low cost, low power consumption and easy deployment \cite{8910627,9475160}.

However, the passive RIS suffers from the multiplicative fading effects, where the propagation link experiences a product of channel attenuation of two individual links.
To address this issue, a novel concept named active RIS was proposed in \cite{zhang2021active,9377648}.
In an active RIS, each reflecting element is comprised of additional reflection-type amplifier, which not only adjusts the phase shift of the incident signals, but amplifies the incident signal power.
The authors of \cite{2021arXiv211207510Z} demonstrated that when both the passive RIS and the active RIS systems are assumed to have the same total power consumption,
the active RIS outperforms the passive RIS in most cases.
In an active RIS-aided single-user single-input single-output system, the authors in \cite{zhang2021active} derived the asymptotic expression for signal-noise-ratio (SNR) when the number of reflection elements approaches infinity.

The existing contributions on the active RIS mostly consider the instantaneous channel state information (CSI).
However, this transmission design scheme requires a high pilot overhead and a high feedback overhead.
To address this issue, a two-timescale CSI scheme is proposed \cite{9198125}.
The beamforming vector at the base station (BS) could be designed on the basis of the instantaneous CSI \cite{9475160}, while
the active RIS configuration is based on the statistical CSI, which only depends on the angle and location information that generally remain invariant in a long term \cite{9656609,8746155}.
In addition, due to the slow varying of the statistical CSI, the computational complexity and feedback overhead can be greatly reduced \cite{9400843}.

Against the above background, we propose a
two-timescale CSI-based transmission design for the active RIS-aided wireless system.
Specifically, we summarize the contributions of this paper in short:
1) we derive a closed-form expression of the achievable rate (AR) of the considered system;
2) we propose a genetic algorithm (GA) to solve the phase shift optimization problem (PSOP);
3) this expression is extended to the downlink system.


\begin{figure*}[b]
\hrulefill
\setcounter{equation}{8}
\begin{equation}\label{rate_zhangqi1}
R_U\approx \log _2\Big( 1+\frac{P_{t,U}\lambda _{U}^{2}\mathbb{E}\left\{ \lVert \mathbf{H}_U\mathbf{\Phi }_U\mathbf{g}_U \rVert _{2}^{4} \right\}}{\lambda _U^2\mathbb{E}\left\{ \lVert \mathbf{g}_{U}^{H}\mathbf{\Phi }_{U}^{H}\mathbf{H}_{U}^{H}\mathbf{H}_U\mathbf{\Phi }_U \rVert _{2}^{2} \right\} \sigma _{V,U}^{2}+\sigma _{N,U}^{2}\mathbb{E}\left\{ \lVert \mathbf{H}_U\mathbf{\Phi }_U\mathbf{g}_U \rVert _{2}^{2} \right\}} \Big)
\end{equation}
\hrulefill
\setcounter{equation}{10}
\begin{equation*}
\mathbb{E}\left\{ |\left[ \mathbf{H}_U\mathbf{\Phi }_U\mathbf{g}_U \right] _m|^2 \right\} =\kappa_UK_1K_2\underset{w_1}{\underbrace{|\left[ \mathbf{a}_M\left( \phi _{r}^{a},\phi _{r}^{e} \right) \right] _mf_U|^2}}+\kappa_UK_1\underset{w_2}{\underbrace{\mathbb{E}\Big\{ \Big| \left[ \mathbf{a}_M\left( \phi _{r}^{a},\phi _{r}^{e} \right) \right] _m\sum_{n=1}^N{\left[ \mathbf{a}_{N}^{*}\left( \varphi _{t}^{a},\varphi _{t}^{e} \right) \right] _ne^{j\theta _{nU}}}\left[ \mathbf{\tilde{g}}_U \right] _n \Big|^2 \Big\} }}
\end{equation*}
\begin{equation}\label{abs}
+\kappa_UK_2\underset{w_3}{\underbrace{\mathbb{E}\Big\{ \Big| \sum_{n=1}^N{\big[ \mathbf{\tilde{H}}_U \big] _{mn}e^{j\theta _{nU}}}\big[ \mathbf{a}_N\left( \varphi _{r}^{a},\varphi _{r}^{e} \right) \big] _n \Big|^2 \Big\} }}+\kappa_U\underset{w_4}{\underbrace{\mathbb{E} \Big\{ \Big| \sum_{n=1}^N{\big[ \mathbf{\tilde{H}}_U \big] _{mn}e^{j\theta _{nU}}}\big[ \mathbf{\tilde{g}}_U \big] _n \Big|^2 \Big\} }}
\end{equation}
\normalsize
\end{figure*}

\section{System Model }

We consider an uplink wireless communication system, where a single-antenna user transmits data information to an $M$-antenna BS via an $N$-element active RIS.
The active RIS can be employed at the facade of buildings, which can provide a strong line-of-sight (LoS) component in wireless links.
Hence, the Rician fading model is considered in the RIS-related channels.
In addition, the channel between the user and the BS is assumed to be blocked by the obstacles.
The uplink channel matrix between the active RIS and the BS is $\mathbf{H}_U\in \mathbb{C} ^{M\times N}$.
The uplink channel vector between the user and the RIS is $\mathbf{g}_U\in \mathbb{C} ^{N\times 1}$.
Specifically, $\mathbf{H}_U$ and $\mathbf{g}_U$ are following Rician fading distribution,
where $\alpha_U$ and $\beta_U$ are the large-scale fading coefficients, $K_1$ and $K_2$ are the Rician factors, which are respectively defined as the power ratio of the LoS components and the NLoS components.
The NLoS components $\widetilde{\mathbf{H}}_U$ and $\widetilde{\mathbf{g}}_U$ consist of independent and identical distribution (i.i.d.) entries, all of which follow the distribution of $\mathcal{CN}(0,1)$.
The LoS components $\overline{\textbf{H}}_U$ and $\overline{\textbf{g}}_U$
apply the uniform
squared planar array (USPA) model, which can be written as
\setcounter{equation}{0}
\begin{equation}
\mathbf{\bar{H}}_U=\mathbf{a}_M\left( \phi _{r}^{a},\phi _{r}^{e} \right) \mathbf{a}_N^H\left( \varphi _{t}^{a},\varphi _{t}^{e} \right) ,
\ \ \ \ \
\mathbf{\bar{g}}_U=\mathbf{a}_N\left( \varphi _{r}^{a},\varphi _{r}^{e} \right) ,
\end{equation}
with
\vspace{-0.25cm}
\begin{equation}
\mathbf{a}_X\left( \vartheta ^a,\vartheta ^e \right) \!\!=\!\!\!\left[\!\!\! \begin{array}{c}
	1,\cdots ,e^{j2\pi \frac{d}{\lambda}\left( x\sin \vartheta ^a\sin \vartheta ^e+y\cos \vartheta ^e \right)},\cdots ,\\
	e^{j2\pi \frac{d}{\lambda}\left( \left( \sqrt{X}-1 \right) \sin \vartheta ^a\sin \vartheta ^e+\left( \sqrt{X}-1 \right) \cos \vartheta ^e \right)}\\
\end{array} \!\!\!\right] ^T\!\!\!\!\!\!,\!\!\!
\end{equation}
where $X\in \left\{ M,N \right\}$, $0\leqslant x,y\leqslant \sqrt{X}-1$.
The angles
$\varphi _{r}^{a}$ and $\varphi _{r}^{e}$ are respectively the azimuth and elevation angles of arrival (A-AoA and E-AoA) at the RIS from the user, the angles $\phi _{r}^{a}$ and $\phi _{r}^{e}$ are the A-AoA and E-AoA at the RIS from the BS, while the angles
$\varphi _{t}^{a}$ and $\varphi _{t}^{e}$ are respectively the azimuth and elevation angles of
departure (A-AoD and E-AoD) from the RIS to the BS.
The parameters $d$ and $\lambda$ are respectively the element spacing and  the carrier wavelength, and we set $d=\frac{\lambda}{2}$ similar as \cite{8910627} to facilitate analysis.

Different from the passive RIS, the active RIS amplifies the incident signals by the external power supply, however the thermal noise is amplified accordingly.
Hence, the uplink signal received at the BS is
\begin{equation}
\mathbf{y}_U=\sqrt{P_{t,U}}\mathbf{H}_U\mathbf{\Lambda }_U\mathbf{\Phi }_U\mathbf{g}_Ux+\mathbf{H}_U\mathbf{\Lambda }_U\mathbf{\Phi }_U\mathbf{v}_U+\mathbf{n}_U,
\end{equation}
where $P_{t,U}$ is the transmission power of the user,
and the scalar $x$ represents its transmitted symbol satisfying $\mathbb{E} \big\{ \left| x \right|^2 \big\} =1$.
The vector $\mathbf{v}_U\in \mathbb{C} ^{N\times 1}\sim {\cal CN}\left( \mathbf{0},\sigma _{V,U}^{2}\mathbf{I}_N \right)$ is the uplink thermal noise at the RIS.
The vector  $\mathbf{n}_U\in \mathbb{C} ^{M\times 1}\sim \mathcal{C} \mathcal{N} \left( \mathbf{0},\sigma _{N,U}^{2}\mathbf{I}_M \right) $ is the additive white Gaussian noise (AWGN) at the BS.
The power amplification factor matrix
$\bm{\Lambda}_U=\mathrm{diag}\left( \lambda_{1U},\lambda_{2U},\cdots ,\lambda_{NU} \right)$ and the phase shift matrix $\mathbf{\Phi}_U=\mathrm{diag}\left( e^{j\theta _{1U}},e^{j\theta _{nU}},\cdots ,e^{j\theta _{NU}} \right)$, where $\theta _{nU}$ is the phase shift of the $n$th RIS element.
For simplicity, the amplification factors for elements in the active RIS are assumed to be identical, i.e., $\lambda_{1U}\!=\!\lambda_{2U}\!=\!\cdots \!=\!\lambda_{NU}\!=\!\lambda_U$.
Then, the amplification factor $\lambda_U$ can be obtained as
\begin{align}\label{eta_U}
\lambda _{U}=&\sqrt{ \frac{P_{r,U}}{\mathbb{E}\left\{ \lVert \sqrt{P_{t,U}}\mathbf{\Phi }_U\mathbf{g}_U \rVert _{2}^{2}+\lVert \mathbf{\Phi }_U\mathbf{v}_U \rVert _{2}^{2} \right\}} }
\nonumber \\
=&\sqrt{ \frac{P_{r,U}}{N\left( P_{t,U}\beta _U+\sigma _{V,U}^{2} \right)} },
\end{align}
where $P_{r,U}$ is the uplink amplification power at the active RIS.

%

In this paper, we adopt a two-timescale transmission design protocol due to its low channel estimation overhead and computational complexity.
The BS uses the maximal ratio combining (MRC) technique to process its received signal.
Therefore, the signal received at the BS can be written as
\setcounter{equation}{4}
\begin{equation}
r_U=\left( \mathbf{H}_U\mathbf{\Lambda }_U\mathbf{\Phi }_U\mathbf{g}_U \right) ^H\mathbf{y}_U.
\end{equation}

Thus, the uplink AR is given by
\begin{equation}\label{rateU}
R_U=\mathbb{E} \left\{ \log _2\left( 1+{\gamma}_U \right) \right\} ,
\end{equation}
where $\gamma _U$ is the SNR defined by
\begin{equation}\label{gamma_U}
\gamma _U\!=\!\frac{P_{t,U}\lambda _U^2\lVert \mathbf{H}_U\mathbf{\Phi }_U\mathbf{g}_U \rVert _{2}^{4}}{\lambda _U^2\lVert \mathbf{g}_{U}^{H}\mathbf{\Phi }_{U}^{H}\mathbf{H}_{U}^{H}\mathbf{H}_U\mathbf{\Phi }_U \rVert _{2}^{2}\sigma _{V,U}^{2}\!+\!\sigma _{N,U}^{2}\lVert \mathbf{H}_U\mathbf{\Phi }_U\mathbf{g}_U \rVert _{2}^{2}}\!.\!\!\!\!
\end{equation}




\begin{figure*}[hb]
\hrulefill
\setcounter{equation}{17}
\begin{equation*}
\xi = \kappa_U^2M^2K_1^2K_2^2|f_U|^4+2\kappa_U^2MK_1K_2|f_U|^2\left( 2MNK_1+MNK_2+MN+2M+NK_2+N+2 \right)
\end{equation*}
\begin{equation*}
\hspace{-1cm}
	+\kappa_U^2MN^2\left( K_2^2+2K_1K_2+2K_1+2K_2+1 \right) +
\kappa_U^2M\left( M+1 \right) N\left( 2K_1+2K_2+1 \right)
\end{equation*}
\begin{equation}\label{power4}
\hspace{-5.3cm}
	+\kappa_U^2M^2N^2\left( 2K_1^2+K_2^2+2K_1K_2+2K_1+2K_2+1 \right)
\end{equation}
\hrulefill
\setcounter{equation}{21}
\begin{equation}\label{wishart}
\nu\!=\!\frac{M^2\alpha _U\kappa _U}{\left( K_1\!+\!1 \right)}\left( K_2\left( N\!+\!2K_1|f_U|^2\!+\!K_{1}^{2}N|f_U|^2 \right) \!+\!N\!+\!2K_1N\!+\!K_{1}^{2}N^2 \right) \!+\!\alpha _U\kappa _UMN\left( K_2N\!+\!K_1K_2|f_U|^2\!+\!N\!+\!K_1N \right)
\end{equation}
\end{figure*}

\section{Achievable Rate Analysis and Phase Shift Design}

In this section, we first derive a closed-form approximate expression for the AR of the uplink system aided by the active RIS.
Then, we compare the AR with that in the passive RIS system under the same total transmission power $P_U$.
Finally, we investigate an AR maximization problem and propose a phase shift design to solve the problem.
\subsection{Achievable Rate Analysis}

\begin{theorem}\label{theorem1}
The uplink AR of the active RIS-aided system can be approximated as
\setcounter{equation}{7}
\begin{equation}\label{R_final}
R_U\!\approx\! \log _2\!\Bigg(\! 1\!+\!\frac{\xi P_{t,U}P_{r,U} }{\!\!\!N\!\left( P_{t,U}\beta _U\!+\!\sigma _{V,U}^{2} \right) \delta\sigma _{N,U}^{2}+\nu P_{r,U}\sigma _{V,U}^{2}} \!\Bigg),\!\!\!
\end{equation}
where $\delta$, $\xi$ and $\nu$ are respectively given by \eqref{power2}, \eqref{power4} and \eqref{wishart} at the bottom of the next page.

\end{theorem}

\begin{IEEEproof}
By using {Lemma 1} in \cite{zhang2014power}, $R_U$ in (\ref{rateU}) can be approximated as \eqref{rate_zhangqi1} at the bottom of this page.
The term $\mathbb{E}\left\{ ||\mathbf{H}_U\mathbf{\Phi }_U\mathbf{g}_U||_{2}^{2} \right\}$ in \eqref{rate_zhangqi1} can be written as
\setcounter{equation}{9}
\begin{equation}\label{power2}
\mathbb{E}\left\{ ||\mathbf{H}_U\mathbf{\Phi }_U\mathbf{g}_U||_{2}^{2} \right\}
\!=\!\sum_{m=1}^M{\mathbb{E}\left\{ |[\mathbf{H}_U\mathbf{\Phi }_U\mathbf{g}_U]_{m}|^2 \right\}}
\triangleq \delta,
\end{equation}
where $[\mathbf{a}]_{m}$ is the $m$th entry of $\mathbf{a}$.
Then, we expand $\mathbb{E}\left\{ |\left[ \mathbf{H}_U\mathbf{\Phi }_U\mathbf{g}_U \right] _m|^2 \right\}$ as \eqref{abs} at the bottom of this page.
In \eqref{abs}, $\kappa_U =\frac{\alpha_U \beta_U}{\left( K_1+1 \right) \left( K_2+1 \right)}$ and $f_U$ is given by
\setcounter{equation}{11}
\begin{equation}\label{fU}
f_U\triangleq \mathbf{a}_{N}^{H}\left( \varphi _{t}^{a},\varphi _{t}^{e} \right) \mathbf{\Phi \bar{g}}_U=\sum_{n=1}^N{e^{j2\pi \frac{d}{\lambda}\left( xk+yq \right) +j\theta _{nU}}\,\,},
\end{equation}
where $ k = \sin \varphi _{r}^{a}\sin \varphi _{r}^{e}-\sin \varphi _{t}^{a}\sin \varphi _{t}^{e}$ and
$q=\cos \varphi _{r}^{e}-\cos \varphi _{t}^{e}$.

To obtain $\mathbb{E}\left\{ |\left[ \mathbf{H}_U\mathbf{\Phi }_U\mathbf{g}_U \right] _m|^2 \right\}$, we calculate $w_2$ in \eqref{abs} for instance
\begin{align}\label{abs2}
w_2&=\mathbb{E}\Big\{ \sum_{n=1}^N{|\left[ \mathbf{a}_{N}^{*}\left( \varphi _{t}^{a},\varphi _{t}^{e} \right) \right] _ne^{j\theta _{nU}}}\left[ \mathbf{\tilde{g}}_U \right] _n|^2 \Big\}
\nonumber \\
&=\mathbb{E}\Big\{ \sum_{n=1}^N{|}\left[ \mathbf{\tilde{g}}_U \right] _n|^2 \Big\} =N.
\end{align}
And the other terms in \eqref{abs} can be obtained similarly as
\begin{align}\label{abs_results}
w_1=|f_U|^2,
w_3 =N,
w_4 =N.
\end{align}
Substituting \eqref{abs2} and \eqref{abs_results} into \eqref{abs}, we have
\begin{align}
\!\!\mathbb{E}\!\left\{\! |\!\left[ \mathbf{H}_U\mathbf{\Phi }_U\mathbf{g}_U \right] _m|^2 \!\right\}
\!\!=\!\!\kappa_U \! \left(\! K_1K_2|f_U|^2 \!\!+\!\! K_1N\!+\!K_2N \!+\! N \!\right)\!\!.\!\!\!
\end{align}
Thus, the term $\delta$ in (\ref{power2}) can be derived as
\begin{equation}
\delta=M \kappa_U \left( K_1K_2|f_U|^2+\left( K_1+K_2+1 \right) N \right).
\end{equation}

Furthermore, the term $\mathbb{E}\left\{ ||\mathbf{H}_U\mathbf{\Phi }_U\mathbf{g}_U||_{2}^{4} \right\}$ in \eqref{rate_zhangqi1} can be written as
\begin{equation*}
\hspace{-1.6cm}
\mathbb{E}\left\{ ||\mathbf{H}_U\mathbf{\Phi }_U\mathbf{g}_U||_{2}^{4} \right\} =
\sum_{m=1}^M{\mathbb{E}\left\{ |\left[ \mathbf{H}_U\mathbf{\Phi }_U\mathbf{g}_U \right] _m|^4 \right\}}
\end{equation*}
\vspace{-0.1cm}
\begin{equation}\label{four1}
+2\!\!\sum_{m=1}^{M-1}{\!\sum_{h=m+1}^M{\!\!\!\!\!\mathbb{E}\left\{ |\left[ \mathbf{H}_U\mathbf{\Phi }_U\mathbf{g}_U \right] _m|^2|\left[ \mathbf{H}_U\mathbf{\Phi }_U\mathbf{g}_U \right] _h|^2 \right\}}}
\!\triangleq\! \xi.\!\!\!\!\!\!
\end{equation}
By using the similar method for calculating \eqref{abs}, we can obtain $\xi$ given by \eqref{power4} at the bottom of the next page.

We then transform the last expectation in \eqref{rate_zhangqi1} as
\setcounter{equation}{18}
\begin{align}\label{ds}
&\mathbb{E}\left\{ \mathbf{g}_{U}^{H}\mathbf{\Phi }_{U}^{H}\mathbf{H}_{U}^{H}\mathbf{H}_U\mathbf{H}_{U}^{H}\mathbf{H}_U\mathbf{\Phi }_U\mathbf{g}_U \right\}
\nonumber \\
&\;\;\;\;\;\;\;\;\;\;\;\;\;\;\;\;\;\;= \mathbb{E}\left\{ \mathbf{g}_{U}^{H}\mathbf{\Phi }_{U}^{H}\mathbb{E}\left\{ \mathbf{W}_U\mathbf{W}_U \right\} \mathbf{\Phi }_U\mathbf{g}_U \right\} \triangleq \nu,
\end{align}
where $\mathbf{W}_U=\mathbf{H}_U^H\mathbf{H}_U$ is a non-central Wishart matrix following $\mathcal{W}\left( M,\mathbf{A}_U,\mathbf{\Sigma }_U \right)$ with $\mathbf{A}_U=\sqrt{\frac{\alpha_U K_1}{K_1+1}}\mathbf{\bar{H}}_U$ and   $\mathbf{\Sigma }_U=\frac{\alpha_U}{K_1+1}\mathbf{I}_N$ \cite{zhang2014power}.
According to \cite{1972Approximations}, the non-central Wishart matrix
$\mathbf{W}_U$
can be approximated by a
central Wishart distribution $\mathcal{W}\left( M,\mathbf{0},\mathbf{\bar{\Sigma}}_U \right)$ with
\begin{equation}
\mathbf{\bar{\Sigma}}_U\!=\!\mathbf{\Sigma }_U\!+\!\frac{1}{M}\mathbf{A}_U^H\mathbf{A}_U\!=\! \frac{\alpha_U}{K_1+1}\Big( \mathbf{I}_N\!+\!\frac{K_1}{M}\mathbf{\bar{H}}_U^H\mathbf{\bar{H}}_U \Big).
\end{equation}
According to \cite{Tague1994}, we obtain $\mathbb{E}\left\{ \mathbf{W}_U\mathbf{W}_U \right\}$ given by
\begin{align}\label{WW}
&\mathbb{E}\left\{ \mathbf{W}_U\mathbf{W}_U \right\}
=\frac{\alpha_U ^2MN}{K_1+1}\Big( \mathbf{I}_N+\frac{K_1}{M}\mathbf{\bar{H}}_U^H\mathbf{\bar{H}}_U \Big)
\nonumber \\
&+\!\frac{M^2\alpha_U ^2}{\!\left(\! K_1\!+\!1 \right) ^2}\!\Big(\! \mathbf{I}_N\!+\!\frac{2K_1}{M}\mathbf{\bar{H}}_U^H\mathbf{\bar{H}}_U\!+\!\frac{K_1^2}{M^2}\mathbf{\bar{H}}_U^H\mathbf{\bar{H}}_U\mathbf{\bar{H}}_U^H\mathbf{\bar{H}}_U \!\Big)\!.\!\!\!
\end{align}
Utilizing \eqref{WW} and the fact that $\mathrm{tr}(\mathbf{AB})=\mathrm{tr}(\mathbf{BA})$, we can further derive $\nu$ as \eqref{wishart} at the bottom of this page.

By substituting \eqref{eta_U}, \eqref{power2}--\eqref{wishart} into \eqref{rate_zhangqi1}, we can obtain \eqref{R_final}.
Thus, the proof is completed.
\end{IEEEproof}

\begin{remark}
We denote $\mathrm{\gamma_{BS}^U}=\frac{P_{r,U}}{\sigma _{N,U}^{2}}$,
and $\mathrm{\gamma_{RIS}^U}=\frac{P_{t,U}}{\sigma _{V,U}^{2}}$.
When $\mathrm{\gamma_{BS}^U}\rightarrow \infty$, the AR in \eqref{R_final} converges to
\setcounter{equation}{22}
\begin{equation}
R_U\rightarrow \log _2\!\Big(1+ \xi\mathrm{\gamma_{RIS}^U}/\nu \Big).
\end{equation}
When $\mathrm{\gamma_{RIS}^U}\rightarrow \infty$, we have
\begin{equation}
R_U \rightarrow \log _2\Big(1+\xi\mathrm{\gamma_{BS}^U}/(N\beta _U\delta)  \Big).
\end{equation}
\end{remark}

In one respect, the equations above show that the SNR $\gamma _U$ keeps a linear relationship with $\mathrm{\gamma_{RIS}^U}$, when $\mathrm{\gamma_{BS}^U}$ is high.
In the other respect, when $\mathrm{\gamma_{RIS}^U}$ is high, the SNR and $\mathrm{\gamma_{BS}^U}$ also keeps a linear relationship.

Then, we further introduce the power consumption of the considered system.
The uplink overall transmission power is denoted by
\begin{equation}\label{P_active}
P_U=P_{t,U}+P_{r,U}+N\left( P_{\text{SW}}+P_{\text{DC}} \right),
\end{equation}
where $P_{\text{DC}}$ is the direct current biasing power for each active RIS element. $P_{\text{SW}}$ is the power consumed by each RIS element to switch the phase shift and control the circuit.
For comparison, we provide the AR of the passive RIS system and its power consumption in the following corollary.
By setting $\lambda_U=1$ and $\sigma _{V,U}^{2}=0$ in \eqref{eta_U} and \eqref{R_final}, we can obtain the AR of uplink passive RIS system in the following corollary.

\begin{corollary}\label{cor-1}
When the passive RIS system has the same overall transmission power $P_U$ as the active RIS system, the AR of the uplink passive RIS system is given by
\begin{equation}\label{cor1}
R_U^P \approx \log _2\Big( 1+\frac{P_{t,U}^\prime \xi}{\sigma _{N,U}^{2}\delta} \Big),
\end{equation}
where $P_{t,U}^\prime$ is the  transmission power at the user in the passive RIS system, and defined as
\begin{equation}\label{P_passive}
P_U=P_{t,U}^\prime+NP_{\text{SW}}.
\end{equation}

\end{corollary}

\begin{figure*}[hb]
\hrulefill
\setcounter{equation}{37}
\begin{equation}\label{zeta3}
\!\tau\!\!=\!\!\frac{M^2\alpha _D\kappa_D}{\!\left(\! K_3\!+\!1 \!\right)}\!\left(\! NK_4\!+\!N\!+\!2K_3K_4|f_D|^2\!+\!2K_3N
\! + \!K_3^2K_4N|f_D|^2\!+\!K_3^2N^2 \right) \!+\!MN\alpha _D\kappa_D\!\left( NK_4\!+\!N\!+\!K_3K_4|f_D|^2\!+\!K_3 N \right)\!\!\!\!\!\!\!\!
\end{equation}
\hrulefill
\setcounter{equation}{45}
\begin{align*}
\mathbb{E}\left\{ \lVert \mathbf{g}_D\mathbf{\Phi }_D\mathbf{H}_D \rVert _{2}^{4} \right\} \! \!=&
	M^2\kappa_D^2K_{3}^{2}K_{4}^{2}|f_D|^4+2M\kappa_D^2K_3K_4|f_D|^2\left( 2MNK_3+MNK_4+MN+2M+NK_4+N+2 \right)
\nonumber\\
	&+M\kappa_D^2N^2\left( K_{4}^{2}+2K_3K_4+2K_3+2K_4+1 \right) +\kappa _D^2M\left( M+1 \right) N\left( 2K_3+2K_4+1 \right)
\end{align*}
\begin{equation}\label{zeta1}
	+M^2\kappa_D^2N^2\left( 2K_{3}^{2}+K_{4}^{2}+2K_3K_4+2K_3+2K_4+1 \right) \triangleq \zeta
\end{equation}
\end{figure*}
\subsection{Phase Shift Design }

We solve the PSOP to maximize the uplink AR by a genetic algorithm (GA)-based method in this subsection.

Specifically, the PSOP is formulated as follows
\setcounter{equation}{27}
\begin{subequations}\label{p}
\begin{alignat}{1}
&\max_{\mathbf{\Phi }_U} \,\, R_U \\
&\mathrm{s}.\mathrm{t}.\,\,\, \theta _{nU} \in \{ 0,2\pi/2^b,\ldots,2\pi(2^b-1)/2^b \}
\nonumber\\
&\ \forall n = 1, \ldots,N.
\end{alignat}
\end{subequations}

Problem \eqref{p} can hardly be solved by conventional optimization methods.
Given the complex mathematical achievable rate expression, the techniques conceived in \cite{2023arXiv230804058D} cannot be directly applied in our scenario.
To cope with the issue, we fall back on the heuristic algorithms.
We solve the above PSOP by tuning the phase shifts with the GA-based approach, which is detailed in \cite[Algorithm 1]{9355404}.
The proposed GA method consists of population initialization, fitness evaluation, selection, crossover, and mutation.
In addition, we take into account the discrete phase shifts case which uses $b$-bit quantification.
Similar as \cite{8269405}, the complexity of the proposed GA algorithm is $N_t*n$, where $N_t$ is the population size, and $n$ is the number of generations evaluated. Moreover, $n$ is determined by the convergence behavior of the GA.

\section{Extension to Downlink Transmission}

In this section, we extend the uplink active RIS-aided communication system to the downlink case.
Based on that, we analyze the data transmission and derive a closed-form approximate expression of the AR.
\vspace{-0.3cm}
\subsection{Downlink System Model }

Similar to the uplink channels, the downlink channels $\mathbf{H}_D$ and $\mathbf{g}_D$ follow Rician fading,
where $\alpha_D$ and $\beta_D$ are the large-scale fading coefficients, $K_3$ and $K_4$ are the Rician factors.
The LoS components, using USPA model, are
$\mathbf{\bar{H}}_D=\mathbf{a}_N\left( \psi _{r}^{a},\psi _{r}^{e} \right) \mathbf{a}_{M}^{H}\left( \omega _{t}^{a},\omega _{t}^{e} \right)$
and
$\mathbf{\bar{g}}_D=\mathbf{a}_N^H\left( \omega _{r}^{a},\omega _{r}^{e} \right)$.
And the NLoS components $\widetilde{\mathbf{H}}_D$ and $\widetilde{\mathbf{g}}_D$ consist of i.i.d. entries, all of which follow the complex Gaussian distribution of $\mathcal{CN}(0,1)$.
The angles $\psi _{r}^{a}$ and $\psi _{r}^{e}$ are respectively A-AoA and E-AoA from the RIS to the BS, the angles $\omega _{t}^{a}$ and $\omega _{t}^{e}$ are respectively A-AoD and E-AoD at the BS to the RIS, while the angles $\omega _{r}^{a}$ and $\omega _{r}^{e}$ are A-AoD and E-AoD at the RIS to the user.

For the downlink transmission, the transmitted signal is first precoded by the BS, then amplified by the RIS.
Thus, the signal received at the user is given by
\begin{equation}\label{rd}
r_D=\mathbf{g}_D\mathbf{\Lambda }_D\mathbf{\Phi }_D\left(\mathbf{H}_D\mathbf{w}_Ds+\mathbf{v}_D \right) +n_D,
\end{equation}
where $s$ represents the transmitted symbol from the BS with $\mathbb{E} \big\{ \left| s \right|^2 \big\} =1$.
The vector $\mathbf{v}_D\in \mathbb{C} ^{N\times 1}\sim {\cal CN}\left( \mathbf{0},\sigma _{V,D}^{2}\mathbf{I}_N \right)$ is the downlink thermal noise at the RIS.
The vector  $\mathbf{n}_D\in \mathbb{C} ^{M\times 1}\sim \mathcal{C} \mathcal{N} \left( \mathbf{0},\sigma _{N,D}^{2}\mathbf{I}_M \right) $ is the AWGN at the user.
The power amplification factor matrix
$\bm{\Lambda}_D=\mathrm{diag}\left( \lambda_{1D},\lambda_{2D},\cdots ,\lambda_{ND} \right)$,
and the phase shift matrix $\mathbf{\Phi}_D=\mathrm{diag}\left( e^{j\theta _{1D}},e^{j\theta _{nD}},\cdots ,e^{j\theta _{ND}} \right)$, where $\theta _{nD}$ is the phase shift of the $n$th RIS element.
We use the maximal ratio transmission at the BS, thus the precoding vector $\mathbf{w}_D$ is
\begin{equation}\label{wd}
\mathbf{w}_D=\sqrt{\mu _D}\left( \mathbf{g}_D\mathbf{\Lambda }_D\mathbf{\Phi }_D\mathbf{H}_D \right) ^H,
\end{equation}
satisfying
\begin{equation}\label{inqe}
\mathbb{E}\left\{ \mathbf{w}_D^H\mathbf{w}_D \right\} = P_{t,D},
\end{equation}
where $P_{t,D}$ is the downlink transmission power, and  the beamforming coefficient $\mu _D$ is
\begin{equation}\label{varpi}
\!\!\mu _D=\frac{P_{t,D}}{\mathbb{E}\left\{ \lVert \mathbf{g}_D\mathbf{\Lambda }_D\mathbf{\Phi}_D\mathbf{H}_D \rVert _{2}^{2} \right\}}.
\end{equation}

Moreover, similar to the uplink system, we assume each RIS element shares the same amplification factor $\lambda _{D}$, i.e., $\lambda_{1D}\!=\!\lambda_{2D}\!=\!\cdots \!=\!\lambda_{ND}\!=\!\lambda_D$.
Then, the amplification factor $\lambda _{D}$  can be derived in the following Lemma.

\begin{lemma}\label{lemma-2}
When the downlink amplification power of the RIS is $P_{r,D}$, the amplification factor $\lambda _{D}$ can be obtained as
\begin{equation}\label{eta_D}
\lambda _D=\sqrt{\rho P_{r,D}/\left( \tau P_{t,D}+\rho N\sigma _{V,D}^{2} \right)},
\end{equation}
where $\tau$ and $\rho$ are respectively given by \eqref{zeta3} (at the bottom of this page) and \eqref{zeta2}.
\end{lemma}

\begin{IEEEproof}
From \eqref{rd} and \eqref{wd}, we can obtain that
\vspace{.1cm}
\begin{align}\label{prD}
P_{r,D}&=\mathbb{E}\left\{ \lVert \mathbf{\Lambda }_D\mathbf{\Phi }_D\mathbf{H}_D\mathbf{w}_D \rVert _{2}^{2} + \lVert \mathbf{\Lambda }_D\mathbf{\Phi }_D\mathbf{v}_D \rVert _{2}^{2} \right\}  \nonumber \\
&=\tau \lambda _{D}^{4}\mu _D  +\lambda_D^2 N \sigma _{V,D}^{2},
\end{align}
\vspace{.1cm}
where $\tau$ is defined as
\vspace{.1cm}
\begin{align}\label{zeta3_2}
 \tau &= \mathbb{E}\left\{ \lVert \mathbf{\Phi }_D\mathbf{H}_D\mathbf{H}_D^H\mathbf{\Phi }_D^H\mathbf{g}_D^H \rVert _{2}^{2} \right\}
\nonumber \\
&= \mathbb{E}\left\{ \mathbf{g}_D\mathbf{\Phi }_D\mathbb{E}\left\{ \mathbf{W}_D\mathbf{W}_D \right\} \mathbf{\Phi }_D^H\mathbf{g}_D^H \right\},
\end{align}
where $\mathbf{W}_D=\mathbf{H}_D\mathbf{H}_D^H$ is a non-central Wishart matrix following $\mathcal{W}\left( M,\mathbf{A}_D,\mathbf{\Sigma }_D \right)$ with $
\mathbf{A}_D=\sqrt{\frac{\alpha _DK_3}{K_3+1}}\mathbf{\bar{H}}_D^H$ and $\mathbf{\Sigma }_D=\frac{\alpha _D}{K_3+1}\mathbf{I}_N$.
The distribution of $\mathbf{W}_D$ can be approximated by $\mathcal{W}\left( M,\mathbf{0},\mathbf{\bar{\Sigma}}_D \right)$, where $\mathbf{\bar{\Sigma}}_D$ is given by
\begin{equation}
\mathbf{\bar{\Sigma}}_D\!=\!\mathbf{\Sigma }_D\!+\!\frac{1}{M}\mathbf{A}_D^H\mathbf{A}_D\!=\!\frac{\alpha _D}{K_3+1}\left( \mathbf{I}_N\!+\!\frac{K_3}{M}\mathbf{\bar{H}}_D\mathbf{\bar{H}}_D^H \right).
\end{equation}
The term $\mathbb{E}\left\{ \mathbf{W}_D\mathbf{W}_D \right\}$ in \eqref{zeta3_2} can be obtained as \cite{Tague1994}
\begin{align}\label{WW_down}
&\!\!\!\mathbb{E}\left\{ \mathbf{W}_D\mathbf{W}_D \right\} =M^2\mathbf{\bar{\Sigma}}_D^2+M {\rm tr}\left( \mathbf{\bar{\Sigma}}_D \right) \mathbf{\bar{\Sigma}}_D
\nonumber \\
&\;\;\;\;\;\;\;=\!\!M^2\!\Big(\! \frac{\alpha _D}{K_3 \!\!+\!\! 1} \!\Big) ^2\Big(\! \mathbf{I}_N \!\!+\!\! 2\frac{K_3}{M}\mathbf{\bar{H}}_D\mathbf{\bar{H}}_D^H \!\!+\!\! \frac{K_3^2}{M^2}\mathbf{\bar{H}}_D\mathbf{\bar{H}}_D^H\mathbf{\bar{H}}_D\mathbf{\bar{H}}_D^H \!\Big)
\nonumber \\
&\;\;\;\;\;\;\;\;\;\; +M\alpha _DN\frac{\alpha _D}{K_3+1}\Big( \mathbf{I}_N+\frac{K_3}{M}\mathbf{\bar{H}}_D\mathbf{\bar{H}}_D^H \Big).
\end{align}
With the aid of \eqref{WW_down}, we can derive $\tau$ as \eqref{zeta3} at the bottom of this page.

Considering the same amplification factor for every element, we can derive the beamforming coefficient $\mu _D$ in \eqref{varpi} as
\setcounter{equation}{38}
\begin{equation}\label{varpi2}
\mu _D=\frac{P_{t,D}}{\rho\lambda _D^2},
\end{equation}
with
\begin{align}\label{zeta2}
\rho &=\mathbb{E}\left\{ \lVert \mathbf{g}_D\mathbf{\Phi}_D\mathbf{H}_D \rVert _{2}^{2} \right\}
\nonumber \\
&=M\kappa_D\left( K_3K_4|f_D|^2+\left( K_3+K_4+1 \right) N \right).
\end{align}
In \eqref{zeta2}, $\kappa_D =\frac{\alpha_D \beta_D}{\left( K_3+1 \right) \left( K_4+1 \right)}$ and $f_D$ is given by
\begin{equation}\label{f_D}
f_D= \mathbf{\bar{g}}_D\mathbf{\Phi }_D\mathbf{a}_N\left( \psi _{r}^{a},\psi _{r}^{e} \right)=\sum_{n=1}^N{e^{j2\pi \frac{d}{\lambda}\left( x \ell+yz \right) +j\theta _{nD}}\,\,},
\end{equation}
where $ \ell = \sin \psi _{r}^{a}\sin \psi _{r}^{e}-\sin \psi _{t}^{a}\sin \psi _{t}^{e}$ and
$z=\cos \psi _{r}^{e}-\cos \psi _{t}^{e}$.

Furthermore, substituting \eqref{zeta3} and \eqref{varpi2} into \eqref{prD}, the final
result can be obtained in \eqref{eta_D}, and the proof is completed.
\end{IEEEproof}

From \eqref{rd} and \eqref{wd}, the downlink AR is given by
\begin{equation}\label{down_rate}
R_D=\mathbb{E} \left\{ \log _2\left( 1+{\gamma}_D \right) \right\} ,
\end{equation}
with
\begin{equation}
\gamma _D=\frac{\mu _D\lambda _{D}^{4}\lVert \mathbf{g}_D\mathbf{\Phi}_D\mathbf{H}_D \rVert _{2}^{4}}{\lambda _{D}^{2}\lVert \mathbf{g}_D\mathbf{\Phi }_D \rVert _{2}^{2}\sigma _{V,D}^{2}+\sigma _{N,D}^{2}}.
\end{equation}

\subsection{Downlink Achievable Rate}
\begin{theorem}\label{theorem2}
The downlink AR can be approximated as
\begin{align}\label{RD_final}
\!\!\!\!R_D\!\approx\!
\log _2 \!\!\left(\!\! 1\!\!+\!\!\frac{\zeta P_{t,D}P_{r,D}}{\rho P_{r,D} \beta _DN\sigma _{V,D}^{2}\!\!+\!\!\left(\! \tau P_{t,D} \!\!+\!\!\rho\sigma _{V,D}^{2}N \!\right) \sigma _{N,D}^{2}} \!\!\right)\!\!\!,\!\!\!
\end{align}
where $\zeta$ is given by \eqref{zeta1} at the bottom of this page.
\end{theorem}
\begin{IEEEproof}
Similar to the uplink system, $R_D$ in (\ref{down_rate}) can be approximated as \begin{equation}\label{RD1}
R_D\approx\log _2\Big( 1+\frac{\mu _D\lambda _{D}^{4}\mathbb{E}\left\{ \lVert \mathbf{g}_D\mathbf{\Phi }_D\mathbf{H}_D \rVert _{2}^{4} \right\}}{\lambda _{D}^{2}\mathbb{E}\left\{ \lVert \mathbf{g}_D\mathbf{\Phi }_D \rVert _{2}^{2} \right\} \sigma _{V,D}^{2}+\sigma _{N,D}^{2}} \Big).
\end{equation}

By utilizing the similar methods in Section III, the term $\mathbb{E}\left\{ \lVert \mathbf{g}_D\mathbf{\Phi }_D\mathbf{H}_D \rVert _{2}^{4} \right\}$ in \eqref{RD1} can be derived as $\zeta$ in \eqref{zeta1} at the bottom of this page. By substituting \eqref{zeta1} and
$
\mathbb{E}\left\{ \lVert \mathbf{g}_D\mathbf{\Phi }_D \rVert _{2}^{2} \right\} =\beta _DN
$
into \eqref{RD1}, we obtain
\setcounter{equation}{46}
\begin{equation}\label{RD2}
R_D\approx \log _2\Big( 1+\frac{\mu _D\lambda _{D}^{4}\zeta}{\lambda _{D}^{2}\beta _DN\sigma _{V,D}^{2}+\sigma _{N,D}^{2}} \Big).
\end{equation}

In addition, substituting \eqref{eta_D}   and \eqref{varpi2} into \eqref{RD2}, we derive the final result in \eqref{RD_final}, and the proof is completed.
\end{IEEEproof}

\begin{remark}
We denote $\mathrm{\gamma_{US}^D}=\frac{P_{r,D}}{\sigma _{N,D}^{2}}$,
and
$\mathrm{\gamma_{RIS}^D}=\frac{P_{t,D}}{\sigma _{V,D}^{2}}$.
When $\mathrm{\gamma_{US}^D}\rightarrow \infty$, the AR in \eqref{RD_final} converges to
\begin{equation}
R_D\rightarrow \log _2\Big( 1+\zeta\mathrm{\gamma_{RIS}^D}/(\rho \beta _DN) \Big).
\end{equation}
When $\mathrm{\gamma_{RIS}^D}\rightarrow \infty$, we have
\begin{equation}
R_D\!\rightarrow\log _2\Big( 1+\zeta\mathrm{\gamma_{US}^D}/\tau \Big).
\end{equation}
\end{remark}

The equations above show similar results as it in the uplink system.

Moreover, we focus on the power consumption of the considered system.
The downlink overall transmission power is denoted by
\begin{equation}\label{P_active}
P_D=P_{t,D}+P_{r,D}+N\left( P_{\text{SW}}+P_{\text{DC}} \right).
\end{equation}
For comparison, the following corollary provides the achievable rate and the power consumption of the passive RIS system.
By setting $\lambda_D=1$ and $\sigma _{V,D}^{2}=0$ in \eqref{prD} and \eqref{RD_final}, we can obtain the achievable rate of the downlink passive RIS system in the following corollary.

\begin{corollary}\label{cor-2}
When the passive RIS system has the same overall transmission power $P_D$ as the active RIS system, the AR of the downlink passive RIS system is given by
\begin{equation}\label{cor2}
R_{D}^{P}\approx \log _2\Big( 1+\frac{\zeta p _D^\prime}{\rho \sigma _{N,D}^{2}} \Big),
\end{equation}
where $P_{t,D}^\prime$ is the transmission power at the BS in the passive RIS system, and defined as
\begin{equation}\label{P_passive}
P_D=P_{t,D}^\prime+NP_{\text{SW}}.
\end{equation}

\end{corollary}

The optimal phase shift for the downlink system can be obtained by using the GA method, which is similar to the uplink system in Section III.
For simplicity, the details are omitted.

\section{Simulation Results}

We analyze the performance of the proposed communication systems,
by setting $\sigma^2_{V,U} =\sigma^2_{V,D} = -70$ dBm, $\sigma^2_{N,U} =\sigma^2_{N,D} = -80$ dBm, $M$ = 128, $N$ = 256, $K_1$ = 10, and $K_2$ = 1, and the large-scale fading coefficients $\alpha$ and $\beta$ are set according to \cite{9355404}.
For convenience, we set $P_{t,U} = P_{r,U}$ and $P_{t,D} = P_{r,D}$ in this paper.

\begin{figure}[t]
\vspace{-1cm}
\centering
\includegraphics[scale=0.55]{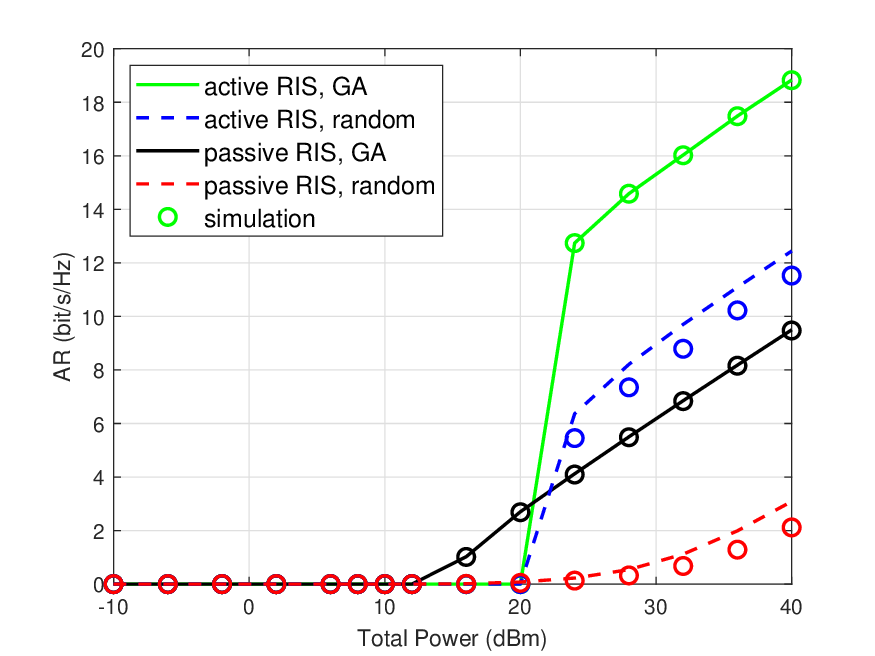}
      \caption{Uplink AR versus $P_U$.} \label{fig1}
\end{figure}

In Fig. \ref{fig1}, we depict the uplink AR versus the transmission power.
The uplink rates are respectively obtained from the approximate analytical expression results in \eqref{R_final} and Monte-Carlo simulation results in \eqref{rateU}.
The former matches the latter well, which demonstrates the accuracy of our derivations.
In addition, the active RIS system outperforms the passive RIS system.
Furthermore, the rate with the optimal phase shifts obtained by the GA-based method performs better than that with random phase shifts, which indicates the importance of optimizing the phase shifts.

\begin{figure}[t]
\vspace{-1cm}
\centering
\includegraphics[scale=0.55]{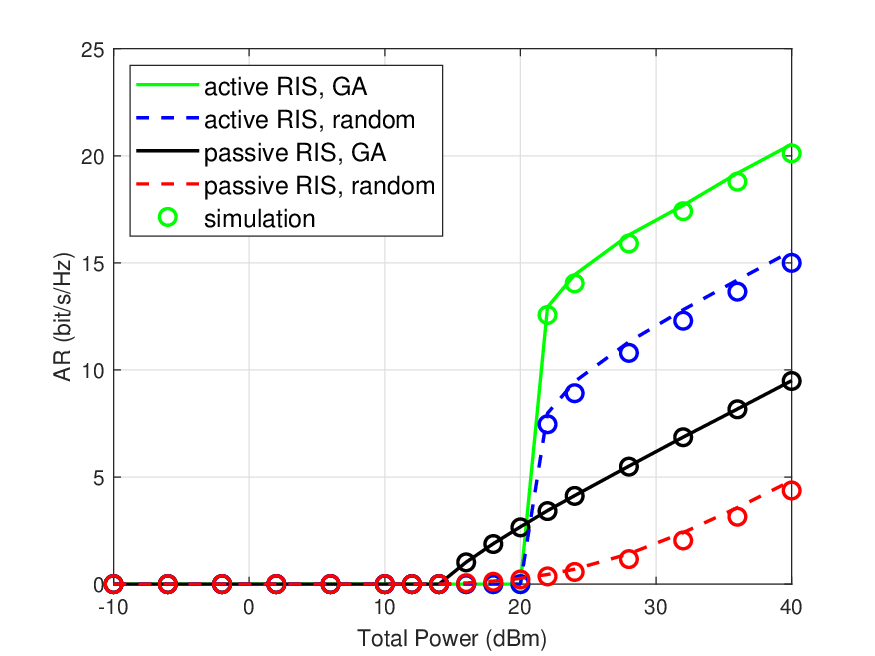}
      \caption{Downlink AR versus $P_D$.} \label{fig2}
\end{figure}

Fig. \ref{fig2} illustrates the downlink AR versus the transmission power.
The uplink rates are respectively obtained from the approximate expression results in \eqref{RD_final} and Monte-Carlo simulation results in \eqref{down_rate}.
It is noticed that when the total power increases from 14 dBm to 20 dBm, the passive RIS system outperforms the active RIS one.
Compared with the passive RIS, the active RIS consumes additional power consumption $P_{\text{DC}}$, and thus its starting power threshold is higher.
However, when the power supply is enough to support the operation of the active RIS, the performance of the active RIS system quickly surpasses that of the passive RIS system.


\section{Conclusion}

This paper investigated an active RIS-aided two-timescale two-way transmission communication.
We derived a tight closed-form approximation for the AR, and the simulation results demonstrated the correctness.
Besides, we adopted the GA method to solve the PSOP of the active RIS.
For comparison, we also simulated the AR achieved by the passive RIS system with both the optimized phase shifts and the random phase shifts.
Simulation results showed that the active RIS system outperforms the passive RIS system when the minimum power requirement of the active RIS is met.

\bibliographystyle{IEEEtran}
\bibliography{IEEEabrv,Refer}

\end{document}